\documentclass[10pt,conference,letterpaper]{IEEEtran}

\usepackage{cite}
\usepackage{booktabs}
\usepackage[hidelinks]{hyperref}
\usepackage{balance}

\usepackage[T1]{fontenc}
\usepackage{makecell}
\usepackage{xspace}
\usepackage{enumitem}
\usepackage{tikz}
\usepackage[normalem]{ulem}
\usepackage{tabularx}

\usepackage{url}

\newcommand\eg{\emph{e.g.},\xspace}
\newcommand\ie{\emph{i.e.},\xspace}
\providecommand{\etal}{\emph{et al.}\xspace}

\newcommand{\crawlingSpan}{12 months\xspace}
\newcommand{\numOfMessages}{70,951,728\xspace}
\newcommand{\numOfMessagesShort}{70.95M\xspace}

\newcommand{\numOfGateways}{29\xspace}
\newcommand{\numOfReceivers}{17,141\xspace}

\newcommand{\numOfCountries}{57\xspace}
\newcommand{\numOfLanguages}{31\xspace}
\newcommand{\numOfServices}{212\xspace}
\newcommand{\numOfMessagesSubset}{46,041,215\xspace}

\begin{document}

\title{Your Code is 0000: An Analysis of the~Disposable~Phone~Numbers~Ecosystem}

\author{
\IEEEauthorblockN{José Miguel Moreno}
\IEEEauthorblockA{\textit{Universidad Carlos III de Madrid} \\
Madrid, Spain\\
josemore@pa.uc3m.es}
\and
\IEEEauthorblockN{Srdjan Matic}
\IEEEauthorblockA{\textit{IMDEA Software Institute} \\
Madrid, Spain \\
srdjan.matic@imdea.org}
\and
\IEEEauthorblockN{Narseo Vallina-Rodriguez}
\IEEEauthorblockA{\textit{IMDEA Networks Institute} \\
Leganés, Spain \\
narseo.vallina@imdea.org}
\and
\IEEEauthorblockN{Juan Tapiador}
\IEEEauthorblockA{\textit{Universidad Carlos III de Madrid} \\
Madrid, Spain \\
jestevez@inf.uc3m.es}
}

\maketitle

\begin{abstract}
  Short Message Service (SMS) is a popular channel for online service
  providers to verify accounts and authenticate users registered to a
  particular service. Specialized applications, called Public SMS
  Gateways (PSGs), offer free Disposable Phone Numbers (DPNs) that can
  be used to receive SMS messages. DPNs allow users to protect their
  privacy when creating online accounts. However, they can also be
  abused for fraudulent activities and to bypass security mechanisms
  like Two-Factor Authentication (2FA). In this paper, we perform a
  large-scale and longitudinal study of the DPN ecosystem by monitoring
  \numOfReceivers unique DPNs in
  \numOfGateways PSGs over the course of \crawlingSpan{}. Using a
  dataset of over 70M messages, we provide an overview of the ecosystem
  and study the different services that offer DPNs and their
  relationships. Next, we build a  framework that (i) identifies and
  classifies the purpose of an SMS; and (ii) accurately attributes every
  message to more than 200 popular Internet services that require SMS for
  creating registered accounts. Our results suggest that the DPN
  ecosystem is globally abused for fraudulent account creation and
  access, affecting all major
  Internet platforms and online services.
\end{abstract}

\section{Introduction}
\label{sec:introduction}

Originating in the late 1990s,
Short Message Service (SMS) have experienced a resurgence among online service providers
(\eg Alphabet, Meta) to deliver notifications, enable Two-Factor Authentication (2FA) and
enhancing the security of online accounts~\cite{google-enforces-2fa,
microsoft-goes-passwordless, twitter-enforces-2fa-employees}.
Online services using SMS-based 2FA technologies
assume that phone numbers are uniquely linked
to an individual. However, this assumption does not hold with
Disposable Phone Numbers (DPNs).
DPNs are shared phone numbers that any individual can use to
receive SMS messages on a public website, so their
metadata and content is published for anyone to see.
Users can take advantage of DPNs to register at online services
without giving their true
personal phone number, either for privacy reasons
or to conduct fraudulent actions.

Despite its potential for abuse, the DPN ecosystem remains relatively
unexplored.
The most recent prior systematic study of DPNs and their usage dates back to
2018~\cite{sending-out-an-sms-2},
a time before the expansion in popularity of SMS-based 2FA
for web and mobile services.
According to Duo Labs, 2FA usage has increased from 28\% in 2017 to 78\% in
2021, becoming the preferred user authentication
method~\cite{duo-labs-auth-report}.
As a result, the key findings of \cite{sending-out-an-sms-2} have
become obsolete.

These reasons, and their potential impact on web services,
motivate us to systematically measure and investigate the
current DPN ecosystem and the evolution of the purposes it supports
in the context of 2FA.
Specifically, we seek answers to the following research questions:
$(i)$ How widely used are DPNs?
$(ii)$ What services are sending messages to DPNs? And
$(iii)$ What is their potential for abuse?
To answer these questions, we develop a methodology to automatically gather
and process a large-scale
and longitudinal
dataset containing \numOfMessagesShort messages received by \numOfReceivers
unique DPNs, collected over a time span close to \crawlingSpan{}.
Using this dataset, we make the following contributions:

\begin{enumerate}[leftmargin=*]
\item[$(i)$] \textit{Study on the usage of DPNs}.
We measure the volume of messages received by DPNs over time.
We find that these numbers receive
collectively more than 1.4M messages per week. A language analysis of
message contents suggest a wide international user base.

\item[$(ii)$] \textit{Service attribution}.
We develop a framework to accurately attribute an SMS message to more than
200 popular global Internet services that require a registered account.
We observe in our dataset messages sent by online service providers of
all sectors, sizes, and geographical scope, including global companies
(\eg Uber, Facebook, Amazon, WhatsApp), security-sensitive industries (\eg
banking), and services developed by smaller and more geographically
localized organizations (\eg Paytm in India).

\item[$(iii)$] \textit{Measuring the potential for abuse}.
We develop a framework to infer and classify the purpose of an SMS
as a proxy to measure their potential for abuse.
We observe that nearly 80\% of messages contain a One-Time Password (OTP),
a single-use link, or both. This figure presents
a significant increase with respect to the trend
reported in the 2018 measurement~\cite{sending-out-an-sms-2}, where this metric was at
67.6\%.
As these messages are closely related to 2FA processes,
we hypothesize that DPN usage is closely related
to anonymity or account fraud.
\end{enumerate}

Our findings suggest that the DPN ecosystem is an expanding and thriving
field, and that the global SMS-based 2FA industry
is oblivious to---or chooses to ignore---potential account abuses arising from it.

\vspace{2mm}
\noindent\textbf{Ethics issues and dataset release}.
The dataset gathered and analyzed in this study might contain sensitive
data since it involves phone numbers that, due to number rotation, might
have belonged in the past or might belong in the future to a real user, and
also SMS messages that can potentially contain personal data or access
credentials. We obtained approval from our IRB to conduct this
study provided that $(i)$ we make no efforts to deanonymize the data in a
way that could facilitate linking messages to actual users; $(ii)$ we
inform affected parties in case that any security or privacy concerns are
identified during the study; $(iii)$ we do not use the collected data for
any secondary purposes other than the scope of this study; and $(iv)$ we
share the dataset on demand with other researchers provided that they agree
on using it for research purposes and under conditions similar to those
described above.
CSV files with the list of analyzed gateways and the services found in the
messages are available at \url{https://github.com/josemmo/your-code-is-0000}.

\section{Background}
\label{sec:background}

\textit{Disposable Phone Numbers} (DPNs) are publicly available phone numbers
offered by \textit{Public SMS Gateways} (PSGs), or simply \textit{gateways}.
DPNs allow receiving messages from a wide catalog of international
phone numbers without the need for a SIM card.
PSGs are usually free services that do not require
registering an account, so multiple users can simultaneously use the same DPN at any time.
However, there are gateways offering ``premium'' DPN services
that require a single-use payment
to read the most recent messages.

The SMS messages received by a DPN are compiled in an \textit{inbox}.
Depending on the gateway offering the service,
some inboxes have a smaller capacity than others. For example,
some only list the latest 30 messages while others index
all messages received over the last months.
SMS messages published in an
inbox typically have four attributes:

\begin{itemize}[leftmargin=*]
\item \textit{Receiver}.
The international number of the message recipient.

\item \textit{Sender}.
The party that, allegedly, sent the message.
Typically, the sender is an online service
(\eg Amazon or WhatsApp) that sends automatically-generated messages.
This data can be displayed as an
international phone number, a short code~\cite{twilio-short-code} or a
sender ID~\cite{twilio-sender-id}.
We note that PSGs providing the sender data in the form of
short code or string sender ID may be incorrect due to their dependency on poorly
implemented or maintained Caller ID Lookups~\cite{bandwidth-caller-id}.

\item \textit{Reception timestamp}.
The date and time when the message was received. It is
displayed as a date string using the timezone of the gateway's
server (\eg ``1st Jan 2022, 12:34 pm'') or as a relative timestamp
(\eg ``12 hours ago''). The latter format allows
accurately determining the date when an SMS message was received but not its time.

\item \textit{Content}.
The actual payload of the SMS, often containing OTP codes and Single-use
Links. OTPs are short numeric codes that are sent to the user
of an online service to verify its identity or confirm an action by inputting
the code in an application. Typically, these codes are 4 to 6 digits long
with dashes or spaces to make them more readable.  Single-use links are the
equivalent to OTPs in the form of URLs, but instead of inputting a code, the
user is expected to click and visit the link.  Some gateways
redact their text content to remove numeric codes.
\end{itemize}

\section{Related work}
\label{sec:related-work}

Reaves \etal conducted in 2016 the first large-scale study of the DPN
ecosystem~\cite{sending-out-an-sms}. Their work
analyzed 400 DPNs in 28 countries and showed
how PSGs
contribute to online account fraud. The authors
conducted a follow up study in 2018 that doubled the size of its dataset but they
did not observe any discernible change in the
ecosystem~\cite{sending-out-an-sms-2}.
Thomas \etal conducted a longitudinal
analysis of Phone-Verified Accounts (PVAs) underground sellers and their infrastructure,
proposing multiple strategies that service operator
can leverage to combat fake accounts~\cite{dialing-back-abuse-on-pva}.
In 2019, Hu \etal measured Disposable Email Services
(DEA)~\cite{disposable-email-services}, which share commonalities with DPN
as both services can be abused for verifying and managing online accounts.
Dmitrienko \etal carried a study on multiple Two-Factor Authentication (2FA)
schemes and weaknesses in their implementation, showing how malware can
intercept messages with single-use codes~\cite{insecurity-of-2fa}. Following
the same research line, Lei \etal investigated how to exploit Android APIs
to steal OTP codes sent through SMS~\cite{insecurity-of-sms-otp}.

\section{Methodology}
\label{sec:methodology}

\begin{figure*}[t!]
   \centering
   \includegraphics[width=\textwidth]{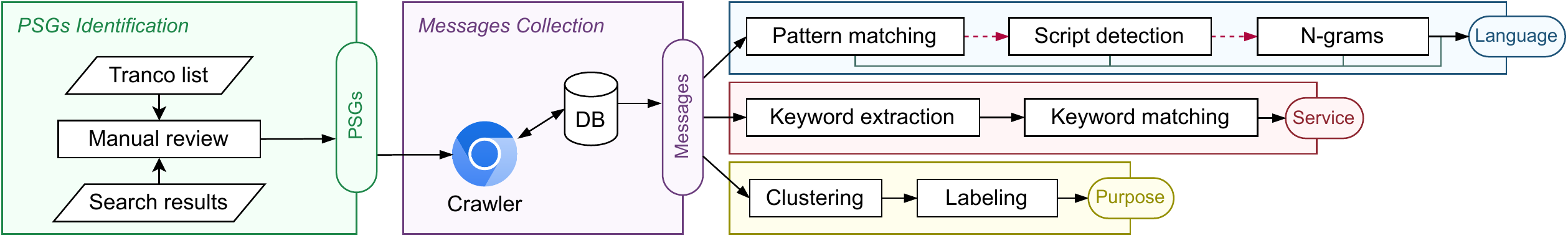}
   \caption{Our pipeline for identifying PSGs, crawling messages and
     post-processing of data. For language detection, arrows with full lines
     (\textcolor{teal}{green}) indicate passed check, while dashed lines in
     (\textcolor{purple}{red}) denote a check that failed.}
   \label{fig:methodology}
 \end{figure*}

This section describes our methodology to identify and crawl the PSGs, and the
post-processing techniques used to parse and analyze the
messages. Figure~\ref{fig:methodology} provides an overview of our pipeline.

\subsection{PSGs Identification}
\label{sec:methodology:identification}
To compile a recent and global list of widely used PSGs, we leverage two
complementary methods:

\begin{enumerate}[leftmargin=*]
\item[$(i)$] \textit{PSGs extraction from the Tranco list}.
Using the Tranco top-3M list\footnote{
  Generated on the 14th of June, 2021.
  Available at \url{https://tranco-list.eu/list/NK2W}.
}~\cite{tranco}, we perform an initial automated token-based search to find
possible PSGs using the Python library WordSegment~\cite{wordsegment} and
keeping the entries matching a set of predetermined keywords.\footnote{
  \textit{sms}, \textit{msg},
  \textit{number},
  \textit{numbers}, \textit{phone},

  \textit{free}, \textit{get}, \textit{receive},
  \textit{online}, \textit{temp}, \textit{otp}, \textit{inbox},
  \textit{virtual}, \textit{verify}, \textit{verification} and \textit{code}.
}
This method identifies 17 sites offering DPN services, 15 of
which are still indexed by Tranco as of Dec. 2022.

\item[$(ii)$] \textit{PSGs extraction from search engines}.
Some gateways are accessed exclusively through mobile apps instead of websites,
so they are likely missed by Tranco. We
leverage Google's search engine to increase our coverage
using the same set of keywords.
This step reveals 12 more gateways, 3 of which are offered by apps
published on Google Play.
\end{enumerate}

We manually review each candidate to discard unrelated, parked or expired domains.
Additionally, we also remove PSGs which $(i)$ are stale
and had not received a single message in months;
$(ii)$ are copies of another gateway
belonging to the same provider; or $(iii)$ are aggregators that
harvest and publish messages from other gateways.
We note that aggregators are easy to identify because they contain duplicate DPNs that
appear in other PSGs, and they publish messages at a slower rate compared to
the original source.

\subsection{Messages Collection}
\label{sec:methodology:crawling}
We use a purpose-built Chromium-based crawler
instrumented with
Playwright~\cite{playwright} to fetch the DPNs and their messages for each
gateway identified in the previous step.
We use an actual web browser instead of a simpler and easier-to-maintain script
because some PSGs need to run JavaScript code on the client-side to properly render
webpages. For those cases, sending crafted HTTP requests and parsing their
responses is not enough.
The PSGs accessible exclusively through Android mobile apps are implemented as
WebViews, and we observe their traffic to find the HTTP requests that fetch the
DPNs from the remote server.
Then, we crawl these gateways with our purpose-built crawler,
sending HTTP requests using the \texttt{fetch} Web API~\cite{mdn-fetch-api}.
This approach allows us not having to maintain two separate codebases.

As mentioned in Section~\ref{sec:background}, some PSGs keep a copy of all
received message for a long time, while others only show the latest $n$
messages. To minimize the number of messages that might get lost, we crawl PSGs
at different sampling periods. We fine-tune the crawler using the reception
timestamp of the oldest message in an inbox and the popularity of the gateway.
We periodically adapt the crawling rate to guarantee that we do not miss too
many messages in case a PSG starts receiving more traffic.
Due to the lack of unique message IDs, we assign our own identifier to messages.
The identifier we use is a composite key formed by the receiver (\ie DPN),
the sender, the reception timestamp and the text content. Thanks to this
identifier, we can track DPNs and messages across different PSGs
to detect and remove duplicates.

\subsection{Language and Service Identification}
\label{sec:methodology:language-and-service}
A DPN can receive messages from online services with a global user base,
and also from other regional ones. For this reason, we first need to detect
the message language.

\vspace{2mm}
\noindent\textbf{Language Detection}.
Guessing the language of short messages is an open research
problem~\cite{apple-very-short-strings}.
We first normalize the message to remove punctuation, duplicate spaces and other
defects (\eg corrupted Unicode characters), and then automatically assign the
language to those messages that contain highly distinctive substrings
(\eg ``verification code'') or use Unicode characters~\cite{unicode-glossary}
unique to world languages (\eg the Modi script in Hindi). If this
process fails, we use the JavaScript franc library~\cite{franc}
to determine the language.

\vspace{2mm}
\noindent\textbf{Sending Service}.
Gateways often include the service name as the ``sender'' from a given message,
but this information might be inaccurate due the presence of services with
outdated Caller ID Lookups. We choose to use our own list of keywords to map a
message to the associated service.
To extract the keywords, we
$(i)$ normalize the message text and remove diacritics using the Normal Form
Decomposition (NFD)~\cite{unicode-normalization-forms},
$(ii)$ tokenize the content separating by whitespace and add the unigrams as
keywords, and
$(iii)$ use the unigrams to generate bigrams and add them as keywords as well.
Then, two of the authors manually analyzed the list of the top-10k most
frequent keywords
and flagged those that were service names or strictly
related (\eg branded domain names, \textit{mottos}).
After this process, we end up with a list of 1.7k meaningful keywords and
\numOfServices unique services.
We note our keyword-based approach might incorrectly attribute a message to a
service if its text contains the service name but it was not sent from the
particular service.
To limit the number of wrongly attributed services, we iterate over the keyword-flagging
process to create more specific keyword-matching rules for the services
with most mislabels.
We measure the accuracy of our service classifier by manually labeling a
sample of 4k randomly selected messages, offering
an accuracy of 99.10\%. Most mislabels are attributed to
false negatives (\ie messages from a known
service but tagged as ``unknown'') in 0.78\% of cases.

\subsection{Purpose Identification}
\label{sec:methodology:purposes}
We use a hierarchical divisive clustering to group message into patterns that
correspond to activities on an account for a given service. In the first
phase,
we group messages depending on the associated service and we normalize the
message content. To generate the normalized version of the message, we
remove multiple whitespaces and punctuation characters, perform stemming, and
filter any stopword or token with less than two characters.
Next, we replace all IBANs, URLs, email addresses, IP addresses, numeric codes
and timestamps found in a message with a fixed pattern.
These patterns or \textit{identifiers} often appear in notification messages
informing of user activity, and can be easily detected using regular expressions
due to their structured format.

We replace each identifier with a pattern that captures the identifier type and
length. For example, an URL with 36 characters is replaced
with ``URL\{36\}'', and a sequence of 4 digits becomes ``NUMERIC\{4\}''. Once this
process is completed, we use the identifiers to group messages into
clusters. Since a message might contain multiple identifiers, our clustering
algorithm prioritizes longer identifiers (\eg a message with ``URL\{36\}'' and
``URL\{20\}'', will be assigned to the cluster ``URL\{36\}'') and those that
better capture the \textit{purpose of the message} (\eg an IP address is
usually more representative than a generic numeric code). For the
priorities, we follow the same order we use to introduce the identifiers.

We tokenize the normalized messages on the whitespaces and then calculate a
fingerprint of the message using SimHash~\cite{www2007manku}.
We leverage the Hamming distance to compare hashes given a
similarity threshold~\cite{pythonsimhash}, which
determines the number of bits that can differ among two
near-duplicates to be considered similar.
We empirically chose a threshold of 10 with 64-bit hashes after
experimenting with different values and validating the results against a
ground truth of 10k randomly-selected normalized messages.

\vspace{2mm}
\noindent\textbf{Lifecycle of an Account}.
Once a user has provided a phone number to a service, the provider can send
messages to notify the user about events associated to their personal account
during their lifecycle. We label messages with a tag that captures the type of
account activity or \textit{purpose} following the NIST SP 800-63B standard.
This specification provides technical guidelines to agencies for the
implementation of digital authentication~\cite{nistpasswordpolicy}. Each one of
the resulting purposes are tied to a particular stage in the lifecycle of
an account:
$(i)$ creation (a new account is created on the service),
$(ii)$ verification (the service requests the user to verify his identity),
$(iii)$ activity (the service notifies of user-performed actions or
important events),
$(iv)$ update (the user's personal data on the service is modified),
and $(v)$ recovery (the service detected an attempt to recover access to an
existing account).

\vspace{2mm}
\noindent\textbf{Automated Message Labeling}.
After defining the categories of purposes to monitor, we randomly
select 6,429 message clusters and manually assign one category to each.
To identify the best label for the cluster, we leverage the normalized
version of the messages with the highest number of occurrences. By
manually inspecting each message, we label the cluster using one of the five
purposes defined above. We also use the message content to
generate patterns that can identify other messages serving a similar purpose.
In the final step, we apply the obtained patterns to automatically infer the
purpose of all messages (and clusters) in our dataset. To evaluate the accuracy
of our automated labeling technique, we randomly select 1k clusters and
pick the content of the normalized message with the highest number of
occurrences to confirm whether the assigned label is correct. We observe that
our approach selects the right label in 91.3\% of the cases being the most common
sources for mislabeling tokens such as
``activity'' (7.8\% of the cases) or ``verification'' (0.32\%).

\section{Ecosystem Characterization}
\label{sec:overview}

\begin{figure*}[t!]
  \centering
  \includegraphics[width=\textwidth]{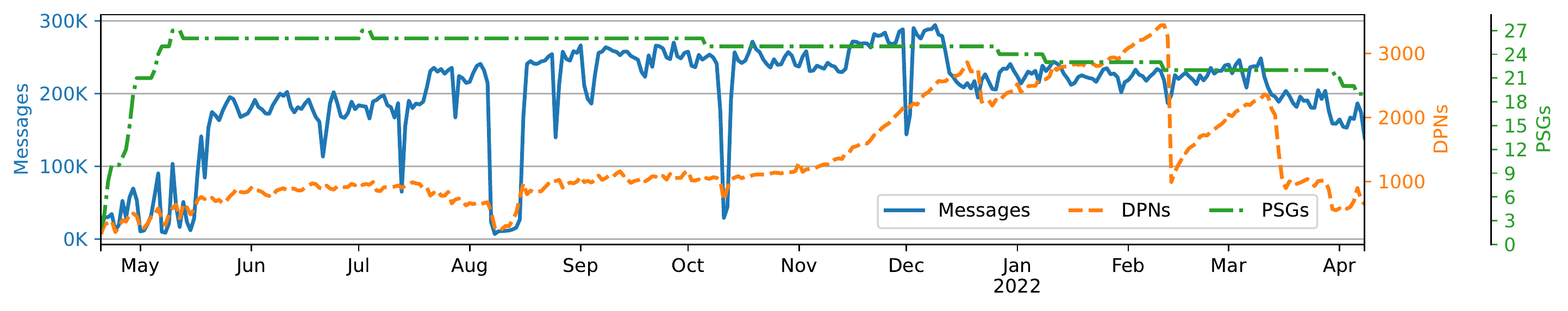}
  \vspace*{-8mm}
  \caption{Daily volume of messages, active DPNs and online PSGs.}
  \label{fig:messages-volume}
\end{figure*}

We collect \numOfMessages messages from \numOfReceivers unique DPNs offered by
\numOfGateways different PSGs over a period close to \crawlingSpan. We note that
the number of gateways decreased over time as some went offline or stopped
publishing new messages.
For reference, the previous measurement from 2018 collected a dataset of 900k
messages across 28 months~\cite{sending-out-an-sms-2}.
Figure~\ref{fig:messages-volume} shows the number of SMS messages, active
DPNs, and online PSGs observed daily. While the weekly count of received
messages is roughly stable, there are two major events on August and
October 2021 where the daily count is considerably lower than on any other
date, and a minor event on December.
These events are unrelated to the DPN ecosystem and were caused by
network downtimes and infrastructure upgrades on our side.

\subsection{Gateways}

\begin{table}[t!]
\centering
\caption{List of crawled PSGs.}
\resizebox{\columnwidth}{!}{
\begin{tabular}{llrrr}
\toprule
Gateway & Crawl start & Days & DPNs & Prefixes \\
\midrule
99dark.com & 2021-05-07 & 336 & 1,151 & 9 \\
bfkdim.com & 2021-04-29 & 7 & 64 & 8 \\
bulk-pva.com & 2021-04-30 & 240 & 3 & 2 \\
cloakmobile.com & 2021-04-29 & 13 & 19 & 3 \\
freebulksmsonline.com & 2021-04-28 & 345 & 501 & 4 \\
freephonenum.com & 2021-04-22 & 351 & 10 & 2 \\
getfreesmsnumber.com & 2021-04-27 & 337 & 907 & 20 \\
onlinesim.io & 2021-04-22 & 351 & 769 & 28 \\
receive-sms-online.com & 2021-04-29 & 66 & 7 & 3 \\
receive-sms-online.info & 2021-04-21 & 352 & 369 & 20 \\
receive-sms.com & 2021-04-28 & 345 & 58 & 2 \\
receive-smss.com & 2021-04-20 & 353 & 420 & 37 \\
receivefreesms.net & 2021-04-22 & 351 & 28 & 3 \\
receivesms.cc & 2021-04-20 & 353 & 56 & 12 \\
receivesms.co & 2021-04-23 & 350 & 41 & 11 \\
receivesms.org & 2021-04-22 & 351 & 16 & 5 \\
sms-online.co & 2021-04-21 & 295 & 5 & 5 \\
sms-receive.com & 2021-07-02 & 95 & 5 & 1 \\
sms-receive.net & 2021-04-23 & 350 & 247 & 17 \\
sms-verification.online & 2021-05-06 & 337 & 22 & 4 \\
sms.sellaite.com & 2021-04-26 & 347 & 12 & 1 \\
sms.visatk.com & 2021-05-05 & 335 & 294 & 5 \\
smsfinders.com & 2021-04-29 & 344 & 6 & 5 \\
smsfree.cc & 2021-04-28 & 345 & 4,733 & 47 \\
temp-sms.org & 2021-05-06 & 247 & 25 & 2 \\
temp99.com & 2021-05-07 & 336 & 8,852 & 13 \\
tempsms.net & 2021-05-10 & 326 & 12 & 2 \\
virtnumber.com & 2021-04-30 & 343 & 66 & 13 \\
www.spoofbox.com & 2021-05-10 & 333 & 8 & 8 \\
\bottomrule
\end{tabular}
}
\label{tab:gateways}
\end{table}

Table~\ref{tab:gateways} lists the PSGs that we crawled during our
research, along with their crawling period (in days), number of DPNs (\ie
inboxes) and number of different calling prefixes.  The number of DPNs
offered by PSGs offer on average 645 DPNs that span across 5 different
calling prefixes. However, we find a high dispersion within these numbers,
with ``bulk-pva.com'' having merely 3 DPNs as opposed to the more than 8k
numbers offered by ``temp99.com''. Both gateways remained active for
hundreds of days.
Over the course of our work, the number of PSGs slowly decreased as some stopped
working properly (\ie receiving new messages), went offline or added protective
measures like CAPTCHAs or JS Challenges to prevent their sites from being crawled.
In this last case, we did not attempt to circumvent such measures for ethical
reasons.

Although we did not perform a longitudinal analysis of the ecosystem, we
repeated the PSG identification process described in Section~\ref{sec:methodology:identification}
18 months after the initial run to find changes in the list of active PSGs.\footnote{
  Using the Tranco list generated on the 11th of December, 2022.
  Available at \url{https://tranco-list.eu/list/JX5LY}.
}
As a result, we find 18 new gateways that did not appear before either because they
did not exist at the time or were not popular enough. In addition, 14 gateways
(half of the entries from Table~\ref{tab:gateways}) are no longer included
in this last outcome.
These changes suggest that many PSGs in the DPN ecosystem are
relatively volatile, with an approximate 1-year lifespan.

\subsection{DPN Dynamics}
In Section~\ref{sec:background}, we hypothesize that a DPN can appear in
different gateways.
We confirm it by noticing that 9.1\% of the DPNs in our dataset appear
in more than one PSGs.
We attribute this behavior to DPN rotation and infrastructure sharing
(\ie DPN reuse).

\vspace{2mm}
\noindent\textbf{DPN rotation}.\xspace
A typical PSG builds its DPN pool by
$(i)$ acquiring and adding to GSM boxes SIM cards from
	multiple mobile network operators; or
$(ii)$ renting VoIP lines from third-party External Short Messaging Entity (ESME) like
       Bandwidth.com~\cite{bandwidthcom} or Twilio~\cite{twilio} to handle the
       reception of messages.
PSGs often distinguishing between so called \textit{``real''}
(mobile lines) and \textit{``virtual''} numbers (VoIP lines) in
their offerings.\footnote{
  The \texttt{receive-sms-online.info} gateway is an example of a website
  advertising numbers ``based on real SIM''.
  On the other hand, \texttt{receive-sms.cc} announces their DPNs as
  ``virtual phone numbers''.
}
VoIP lines can be rented and canceled
at any time, either manually or programmatically.
The number
can eventually be rented by another provider.

\vspace{2mm}
\noindent\textbf{DPN reuse}.\xspace
Some PSGs may share a considerable amount of DPNs that cannot be explained
by DPN rotation. We hypothesize
that a given operator might reuse the same infrastructure across
different gateways under their management or ownership.
A clear case of this is the pair formed by \texttt{99dark.com} and
\texttt{temp99.com}, where the former is the API endpoint for an Android
app~\cite{com.tsoft.moshnumbers} and the latter is a public website.
Besides sharing 483 DPNs, other reasons suggest that these two
gateways are operated by the same group, such as similar naming and
contact information, and same DNS nameservers from Cloudflare.
Another not-so-straightforward case is the \texttt{smsfree.cc} gateway, which
has hundreds of DPNs in common with other seemingly unrelated PSGs, and is
used by (at least) a website~\cite{receive-sms.cc} and an
Android app~\cite{com.receivesms.online}.
The simplest explanation is that \texttt{smsfree.cc} is a white-label API
endpoint reused by many platforms (both web and mobile ones) that lends
its DPNs to other parties as if it were a mobile carrier.
However, given that this PSG also rotates DPNs very frequently (sometimes even
daily), it can also be the case that it aggregates phone numbers from other gateways
on purpose as soon as they become available to rent.

\vspace{2mm}
\noindent\textbf{DPN lifetime}.\xspace
Table~\ref{tab:dpn-lifetime} reports the lifetime (in hours or days) during which
a DPN is active. To determine the activity window, we calculate the amount of
time that passed between the first and the last time when a SMS is sent to a
particular DPN. We identify three main patterns: short lived DPNs (12\%), phone
numbers that are active up to four weeks (58\%), and DPNs that receive messages
for several months (29\%). When inspecting the volume of messages that a phone
number receives, we find that it is usually proportional to the DPN lifetime.
We observe a similar trend both in the subset of messages sent by a service known
by our classifier (``Msgs. w/ service'' row in Table~\ref{tab:dpn-lifetime}) and
in the entire dataset.
Long-lasting DPNs tend to receive more messages, being responsible
for over 81\% of the total messages that we collected. The scenario changes when we
check the total number of services for which we observe at least one message. In
this case, the breakdown depending on the DPN activity is smoother, and
phone numbers that were active for over one month receive messages from only
11\% additional services when compared to DPNs which are active for less than
one day.

\begin{table}[t!]
\centering
\caption{
  DPNs lifetime.
  Values for the mean and the quartiles (Q) are expressed in \textit{days} except for
  entries marked with an asterisk (*), where the unit is the \textit{hour}.
}
\begin{tabular}{lrrrrr}
\toprule
                 & \textless 1d* & 1d - 1w & 1w - 1m & \textgreater 1m & Any    \\
\midrule
Mean             & 8.04          & 3.61    & 18.30   & 149.77          & 51.71  \\
Q1               & 2.89          & 1.86    & 11.27   & 42.36           & 5.05   \\
Q2               & 6.17          & 3.35    & 18.03   & 70.37           & 17.59  \\
Q3               & 11.56         & 5.32    & 26.42   & 134.16          & 35.53  \\
DPNs             & 2,097         & 2,943   & 6,943   & 4,908           & 16,891 \\
Messages         & 279k          & 2.1M    & 10.7M   & 57.8M           & 70.9M  \\
Msgs. w/ service & 96k           & 798k    & 4.3M    & 32.1M           & 37.3M  \\
Services         & 188           & 199     & 211     & 212             & 212    \\
\bottomrule
\end{tabular}
\label{tab:dpn-lifetime}
\end{table}

The lifetime distributions that we observe have some implications.
First, we notice services being associated with DPNs as soon as the phone number
becomes available in a gateway. This might be a possible indicator of abuse and
we explore it further in Section~\ref{sec:analysis:abuse}.
Second, DPNs with a long lifetime unnecessarily expose users to higher privacy
and security risks. This is the case for services that leak Personally
Identifiable Information (PII) such as email addresses or usernames in the SMS they
send. While some gateways only display the messages received in the
last 24 hours (see Section~\ref{sec:background}), others provide access to
historical information.
The lifetime of a DPN does not only increase the time window in which scammers
can collect sensitive information, but it exposes users to other
risks. An attacker that harvests login information by scraping DPN messages
can attempt to perform an account takeover if the DPN is still active and is
used as a recovery mechanism for accessing an account. Alternatively, the DPN can
be abused to discover previously unknown user credentials, in case a service uses
SMS messages during the procedure of recovering forgotten usernames and email
addresses.
Lastly, we notice that if we merge into a single class the two sets of short-lived
DPNs (\ie those that live up to one week), the resulting set of 5,040 numbers received at least one
message from all the services that we monitor. This suggests that the \numOfServices
services that we monitor are extremely popular and widely used across different
countries.

\begin{figure*}[t!]
  \centering
  \includegraphics[width=\textwidth]{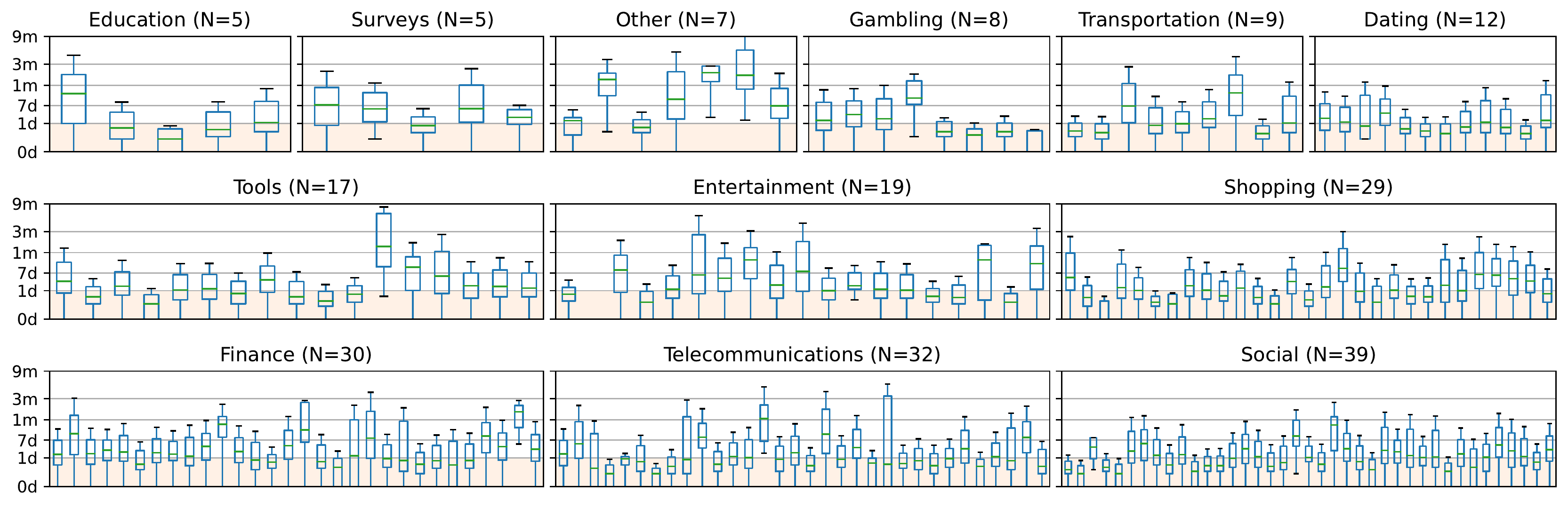}
  \vspace*{-7mm}
  \caption{Time-to-First-Message per service grouped by service category.}
  \label{fig:ttfm-per-service}
\end{figure*}

\subsection{Country and Language Diversity}
The international scope of our analysis requires us dealing with DPNs receiving
messages in any language. We use the international phone number
prefix of a DPN to associate it to a country (\eg ``+44'' is
the international code for the United Kingdom). This allows us to locate DPNs in
\numOfCountries different countries.
Using the language detection method described in
Section~\ref{sec:methodology:language-and-service}, we identify up to \numOfLanguages
languages in our dataset.
The diversity of languages we find suggests an international demand for DPN
services.
With our approach, we could not conclude the language in merely 4.9\% of the messages
either due to their short length and the presence of ambiguous words
(\eg ``Code: 0000'').
Unsurprisingly, English is the most prevalent language in our dataset, accounting
for over 75\% of the messages, followed by a long tail of other languages such as
Indonesian (3.0\%),
French (2.9\%),
Portuguese (2.8\%),
Spanish (2.1\%),
Arabic (1.7\%),
Chinese (1.6\%),
and Russian (1.0\%).
We do not find a 1-to-1 mapping of languages and countries where this language
is spoken. For example, only 30\% of German-written messages are sent to DPNs with the
German international call prefix.
Instead, we see messages being sent globally regardless of their
language.
This global scope is clearly appreciable for English
messages, half of which are sent to non-English
speaking countries.

\subsection{OTPs and Single-use Links}
In their 2018 study, Reaves et al.~\cite{sending-out-an-sms-2} found that 67.6\% of the
messages sent to DPNs contained a code or OTP, thus concluding that receivers were
being used for account verification and user authentication~\cite{sending-out-an-sms}.
In this paper, we extend this methodology and distinguish between
Single-use Codes (\ie OTPs) and Single-use Links. We find OTPs are still on the
rise, with 77.02\% of messages containing them. Single-use Links are less popular,
being present in just 2.18\% of messages, followed by 0.80\% offering both an OTP and
a link.
Given that only 14M messages in our dataset (20\%) have neither of these
single-use means, and that sending services use them to verify authenticated actions
(\ie that require user intervention), we conclude that DPNs are
predominantly being used to create accounts on online platforms.

\subsection{Malicious URLs}
After discarding the aforementioned Single-use Links, we end up with 451,165 messages
that contain a URL. These amount to 178k unique URLs after removing duplicates.
In an attempt to find malicious or harmful URLs in messages sent to DPNs, we use the
Google Safe Browsing API~\cite{google-safe-browsing} to identify web resources
flagged as phishing, malware or spam.
To account for shortened links, we expand their URLs before checking them against
Safe Browsing's database. This expansion is performed by sending an HTTP request to
the shortened URL and retrieving the final ``location'' header without
effectively loading the contents of the destination website.
We use a list of publicly-known shortener services to determine which items need to
be expanded.\footnote{
  See \url{https://github.com/boutetnico/url-shorteners}.
}

With this pipeline, we find merely 41 URLs spread across 125 messages that are considered
harmful by Google. All of them fail in the ``social engineering'' category.
Most malicious URLs appear to be either Apple-related scams
or phishing campaigns targeting banks.

\section{Analysis of Services}
\label{sec:analysis}

This section presents the analysis of the services found sending SMS to
DPNs. Our analysis
focuses on answering two questions: $(i)$ whether DPNs are actually being used for creating
accounts on online services, and $(ii)$ measuring their potential for abuse.
Given the technical limitations of qualitatively analyzing all 70M messages in
our dataset (which includes a long tail of small and lesser known services),
we focus on the subset of \numOfMessagesSubset entries only containing SMS from
the top \numOfServices services. We still consider it a representative subset of
the ecosystem as it is an order of magnitude larger than the whole dataset from
the previous study~\cite{sending-out-an-sms-2}.
As mentioned in Section~\ref{sec:methodology:language-and-service}, this list
of services is based on the most occurring keywords found in our dataset.

Table~\ref{tab:services} lists the top-10 online sending services with more messages
to DPNs.
We find lesser-known ---yet demanded--- services like
DENT~\cite{dent} (which offers free mobile phone lines), regional
operators like Disney+ Hotstar~\cite{hotstar}
(an Indian streaming platform recently acquired by The Walt Disney Company),
and well-known companies with a global userbase (\eg Uber, WhatsApp).
Services like Google, Netflix, Telegram and Tinder
are not included in Table~\ref{tab:services}
because, individually, they account for less than 1.5\% of the total number of messages.
In fact, there is a long tail of sending services in the DPN ecosystem.
We find examples of companies operating in a particular country or world
region (\eg Smood~\cite{com.jamtech.marabel.smood} and
Careem~\cite{careem}, a mobility app from Uber used in the
Middle East) and recognized names in the
Finance (\eg HSBC, Chase, Citibank),
Telecommunications (\eg AT\&T, Deutsche Telekom, Overbit) and even
Public Administration (\eg NHS, government agencies from Spain and India) sectors.

\begin{table}[t!]
\centering
\caption{Top services by volume of messages.}
\begin{tabular}{llrr}
\toprule
Service & Category & Messages & DPNs \\
\midrule
Uber             &  Transportation  &  10.2\%  &   58.2\%  \\
DENT             &  Communications  &   8.4\%  &    4.1\%  \\
Facebook         &  Social          &   3.7\%  &   77.2\%  \\
WhatsApp         &  Communications  &   2.8\%  &   69.3\%  \\
Instagram        &  Social          &   2.8\%  &   68.2\%  \\
PayPal           &  Finance         &   2.6\%  &   45.0\%  \\
Amazon           &  Shopping        &   2.6\%  &   69.3\%  \\
Shopee           &  Shopping        &   2.5\%  &   12.1\%  \\
Disney+ Hotstar  &  Entertainment   &   2.5\%  &  0.292\%  \\
TikTok           &  Social          &   1.5\%  &   72.5\%  \\
\midrule
Other            &  N/A             &  35.1\%  &   99.8\%  \\
\bottomrule
\end{tabular}
\label{tab:services}
\end{table}

Overall, we find the top services by volume account for 65\% of all messages
received by DPNs and offer an ample range of services, including Social Media,
Entertainment, Education, and even sensitive ones such as Telecommunications
and Finance.
Globally-known services like TitTok, Facebook, WhatsApp and Amazon
 have their messages spread across more than
half of all DPNs.
Conversely, highly localized services like the aforementioned Disney+ Hotstar
have a higher density of messages that concentrate in just 50 DPNs,
41 of which have an Indian country prefix.
DENT is an interesting case, as it is the second service with most messages sent to
DPNs yet it only appears in 694 receivers from 45 different countries.
If we sort by DPN coverage, we find two online services with more than 10k receivers
that do not appear in Table~\ref{tab:services} as they each account for less than
0.50\% of messages. These services are Bigo Live~\cite{sg.bigo.live} and
Kwai~\cite{com.kwai.video}, two social media platforms similar to
TitTok functionality-wise.
The median number of services that appear in a DPN is 40, with a maximum of 167
services (covering 78.7\% of the known services lists). All the DPNs we found
in our dataset contain at least one service from the top \numOfServices.

\subsection{Evidence of Usage}
\label{sec:analysis:usage}

Given all the online services we labelled require users to create an account, we
can safely assume these messages are sent by the service as a consequence of an
action performed by the user (\eg registration, login, transaction confirmation).
Therefore, users must be registering accounts on online services using DPNs in order
for these messages to appear in our dataset.
Besides some exceptions like Google, most services make a 1-to-1 relation
between user accounts and phone numbers. In practice, this means that a
given phone number can only be tied to a single service account. For this
reason, and considering that DPNs are by definition shared between multiple
users, there is an incentive to register an account on a popular service as
soon as a DPN becomes available on a gateway.

To verify this dynamic, we measure how long it takes for an online service
to appear in a DPN, presumably as a direct result of an account registration or
recovery (\ie re-verification) event.
Figure~\ref{fig:ttfm-per-service} provides boxplots for each service grouped
by category.
These boxplots represent the \textit{Time-to-First-Message} (TTFM) since the
DPN was first seen in a gateway until a message from a given service was received.
We find a considerable amount of services, especially in the telecommunications
category, have a median TTFM shorter than 24 hours (colored area in the chart).

\subsection{Potential for Abuse}
\label{sec:analysis:abuse}

\begin{figure*}[t!]
  \centering
  \includegraphics[width=\textwidth]{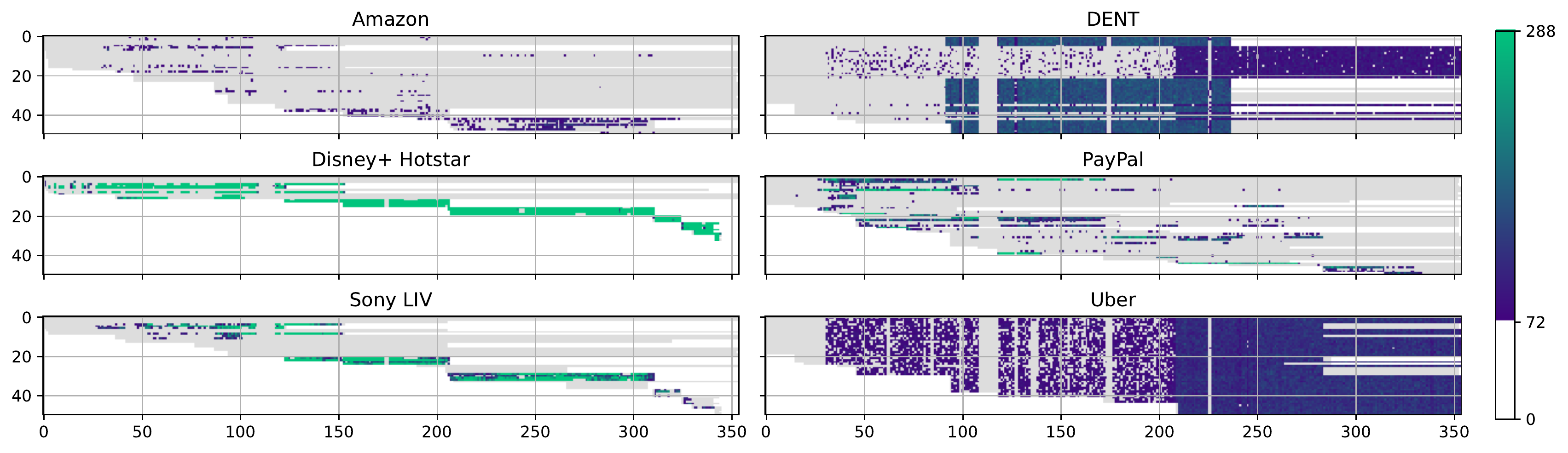}
  \vspace*{-7mm}
  \caption{
    DPNs with the longest bursts of account-related messages grouped by service.
    X axis shows days, while every row in the Y axis plots the messages for a given DPN.
    \textcolor{gray}{Gray} pixels show the lifespan of a DPN.
  }
  \label{fig:abuse}
\end{figure*}

We also analyze the rate at which messages from a specific service are
being sent to a given DPN. As discussed in
Section~\ref{sec:analysis:usage}, for a service message to appear in a DPN,
a user must have previously performed an action that triggered the
message.  Because service messages involve manual interaction, too many
messages in a short period of time might be indicative of some sort of
automation.  While we cannot conclude that the presence of automation is
always related to service abuse (\eg creation of bot accounts, fake
engagement~\cite{fake-engagement}), it is definitely abnormal for a
legitimate user to request an OTP verification code or similar multiple
times per hour on the same account and for an extended time.

Figure~\ref{fig:abuse} shows the services for which we detect
long bursts of account-related messages (\ie those having any of the
purposes mentioned in Section~\ref{sec:methodology:purposes} except
``activity''). For each subplot, we pick the top 50
DPNs with the most received messages and only color the days when a phone
number received at least 72 messages (roughly equivalent to 3 messages per
hour). We also represent as continuous gray lines the DPN lifespan.
These services contain DPNs with
bursts that extend several days, meaning that some phone numbers kept
receiving more than 72 messages per day for more than 100 days straight. In
the case of Disney+ Hotstar and Sony LIV, we can attribute this behavior to
the two services being only available in India and to the scarcity of DPNs
with an Indian calling prefix.  For DENT and Uber, we observe this dynamic
across a much larger pool of phones.

While there are legitimate uses for DPNs (\eg registering anonymously on
dating sites for privacy reasons), DPNs can also be abused to create fake
accounts either manually or using some automation tool as
Figure~\ref{fig:abuse} suggests.  In fact, many PSGs
have banners promoting or advertising services offering PVAs
(\ie phone-verified online accounts available for sale
on underground sites, mostly for nefarious
purposes like fake social engagement~\cite{dialing-back-abuse-on-pva}). We
posit that a reasonable explanation for that is that the PVA provider and
the PSG is the same organization or they are affiliated in some way.

\section{Case Studies}
\label{sec:case-studies}

We next present several cases of interest found in our dataset that
illustrate the potential for abuse of DPNs.

\vspace{2mm}
\noindent\textbf{Free Trials}.\xspace
Oftentimes, popular services offer promotions
or \textit{trials}
to new users. DPNs allow users to register multiple accounts and obtain access
to features that would not be available beyond the trial period.
For example, the Indian online e-learning platform BYJU offered a free 1-to-1
class to newly registered users~\cite{com.byjus.thelearningapp}.  Our dataset
contains around 15k messages from this platform, 7\% of which are associated
with activities, when the service notifies of user-performed actions or
important events. The remaining 92.5\% are either associated with ``account
creation'' or ``account verification'', suggesting that users leverage DPNs to
test the service or access features that would not be available without paying a
subscription.

\vspace{2mm}
\noindent\textbf{Phone Chaining}.\xspace
One of the most interesting uses we see for DPNs is the registration of
private secondary phone numbers.
This is a common practice in the PVA ecosystem and it is usually known
as \textit{phone chaining}~\cite{dialing-back-abuse-on-pva}.
Some big actors enabling this are DENT~\cite{dent} (8\% of the dataset are
messages from this service), Google Voice~\cite{google-voice} and
TextNow~\cite{textnow}.
In fact, phone chaining is so common in this ecosystem that more than 6K DPNs
(36\% of the dataset) have been, at some point, registered in one or more
of just the previous three services.
We also looked for popular mobile phone carriers outside the list of top
services and found evidence of new lines being registered online and then
their SIM cards being sent
by mail. We have obtained evidence of such practices
even for carriers located in countries with
registration laws mandating Proof of Identity, such as
Australia, India and France~\cite{gsma-sim-survey}.

\vspace{2mm}
\noindent\textbf{Finance}.\xspace
In this category we find cryptocurrency
exchanges, FinTech (Financial Technology) apps, and traditional banks.
In all cases, we see messages denoting successfully completed account
creation and transactions.
One of such examples is Empower, a FinTech app that offers microloans. This
service has sent messages to at least 131 different DPNs in multiple occasions
confirming the deposit of funds, meaning that it can potentially be abused for
loan fraud.
Outside the top \numOfServices services list,
we find users linking DPNs to bank accounts to receive verification
codes (\eg Citibank, HSBC, Barclays) and even opening entirely new
accounts, raising concerns for whether this complies with Europe's
PSD2 Strong Customer Authentication (SCA) requirement~\cite{psd2-sca}.
We found more than 100 DPNs registered against
banks, although this figure is probably a lower bound estimate
as we have not thoroughly
looked at this matter nor is the focus of this study.

\vspace{2mm}
\noindent\textbf{Healthcare}.\xspace
We find evidence of DPNs being used for registering to
medical services such as the
British National Health Service (NHS)~\cite{com.nhs.online.nhsonline}
or CoWIN (India's COVID-19 Vaccination Program)~\cite{cowin}.
In both cases we find sensitive information being sent over SMS after the user has
registered to the service. This includes secret single-use codes, COVID-19 test
results and appointments including their precise date and location. We also
identify cases where the names and surnames of the user are sent alongside the
previous information.

\vspace{2mm}
\noindent\textbf{Public Administrations}.\xspace
We find various types of
SMS messages sent by government agencies. While in most cases these are
just innocuous notifications, there are two cases that raise our attention.
``Cl@ve''~\cite{clave-idp} is an Identity Provider used to
authenticate against the Spanish Public Administration by sending an OTP code
to a registered phone number with every login attempt. The presence of this
service in the dataset is concerning since it is tied to a citizen or
a registered company, and can be used to perform sensitive procedures.
We also encounter messages related to ``Aadhaar,'' India's ID
system and the largest in the world~\cite{aadhaar}, which also sends
OTP codes when logging in to banks and other online
services~\cite{aadhaar-faq}.

\section{Conclusions}
\label{sec:conclusions}

Online services have recently doubled down on their efforts to implement
account verification and 2FA flows, using SMS as one of the channels to
deliver these messages.
In this paper, we show that the DPN ecosystem is mostly being (ab)used for
circumventing these security mechanisms
without needing a personal phone number.
We also observe a significant increase in the usage of DPNs
for creating fake accounts since the last
available measurement from 2018, jumping from thousands of messages
received per year to millions.
We find that operators of the PSGs offering DPNs seem to be, in some
cases, strongly tied to underground markets offering
Phone Verified Accounts (PVAs).
We also find that online services do not have effective protections against
the abuses of this ecosystem: Both global well-recognized services (such as
Google and Facebook), banks, governments, and small brands send verification
messages to DPNs.

\vspace{2mm}
\noindent\textbf{Future Work}.\xspace
Our keyword-based message classification shows a huge long tail of services and
purposes yet to be analyzed. Given the scale of the dataset, a more in-depth
automated analysis is needed to understand what trends lie in this long tail.
One interesting research challenge is the definition of methodologies to
identify DPNs and create effective blocklists to mitigate their potential for
abuse when accounts are created.
A second aspect we plan to explore in our future work is a detailed analysis of
the messages with OTP tokens, investigating both message senders and
recipients. Across our dataset, we observe nearly 80\% of the messages with an
OTP code. Fraudsters often abuse these messages to artificially inflate traffic
to a range of numbers controlled by a single mobile network operator; in return,
the mobile operator shares with the fraudster a portion of the generated
revenue~\cite{twilo-sms-traffic-pumping-fraud}.

\section*{Acknowledgements}
This research was supported by
the AEI grant ODIO (PID2019-111429RB-C21 and PID2019-111429RB-C22) and
the Region of Madrid grant CYNAMON-CM (P2018/TCS-4566), co-financed by
European Structural Funds ESF and FEDER.
José Miguel Moreno was supported by the Spanish Ministry of Science and Innovation
with a FPI Predoctoral Grant (PRE2020-094224).
Srdjan Matic was partially supported by the Atracción de Talento grant (Ref. 2020-T2/TIC-20184), funded by Madrid regional government,
and the PRODIGY Project (TED2021-132464B-I00), funded by MCIN/AEI/10.13039/501100011033/ and the European Union NextGenerationEU.
Narseo Vallina-Rodriguez was supported by a Ramon y Cajal Fellowship (RYC2020-030316-I).

The opinions, findings, and conclusions, or recommendations expressed are those
of the authors and do not necessarily reflect the views of any of the funding bodies.

\balance
\bibliographystyle{plain}
\bibliography{paper}

\begin{thebibliography}{10}

\bibitem{receive-sms.cc}
{Receive SMS Online}.
\newblock \url{https://receive-sms.cc/}, 2022.

\bibitem{bandwidth-caller-id}
{Bandwidth Inc.}
\newblock {Caller ID Lookup/Caller Name Lookup}.
\newblock \url{https://www.bandwidth.com/glossary/caller-id-lookup/}, 2022.

\bibitem{bandwidthcom}
{Bandwidth.com}.
\newblock {Bandwidth - The universal platform for global enterprise
  communications}.
\newblock \url{https://www.bandwidth.com/}, 2022.

\bibitem{sg.bigo.live}
{Bigo Technology Pte. Ltd.}
\newblock {Bigo Live – Live Stream, Go Live}.
\newblock \url{https://play.google.com/store/apps/details?id=sg.bigo.live},
  2022.

\bibitem{apple-very-short-strings}
Bishal Bishal and Jerome~R. Bellegarda.
\newblock {Language Identification from Very Short Strings}.
\newblock
  \url{https://machinelearning.apple.com/research/language-identification-from-very-short-strings},
  2019.

\bibitem{com.byjus.thelearningapp}
{BYJU'S}.
\newblock {BYJU'S – The Learning App}.
\newblock
  \url{https://play.google.com/store/apps/details?id=com.byjus.thelearningapp},
  2022.

\bibitem{careem}
{Careem Networks FZ-LLC}.
\newblock Careem.
\newblock \url{https://www.careem.com/}, 2022.

\bibitem{duo-labs-auth-report}
Chrysta Cherrie.
\newblock {The 2021 State of the Auth Report: 2FA Climbs, While Password
  Managers and Biometrics Trend}.
\newblock
  \url{https://duo.com/blog/the-2021-state-of-the-auth-report-2fa-climbs-password-managers-biometrics-trend},
  2021.

\bibitem{twilio-sender-id}
Billy Chia.
\newblock {What is an Alphanumeric Sender ID?}
\newblock
  \url{https://www.twilio.com/docs/glossary/what-alphanumeric-sender-id}, 2022.

\bibitem{dent}
{DENT Wireless Limited}.
\newblock {DENT – The world's first digital \& global operator}.
\newblock \url{https://www.dentwireless.com/}, 2022.

\bibitem{insecurity-of-2fa}
Alexandra Dmitrienko, Christopher Liebchen, Christian Rossow, and Ahmad-Reza
  Sadeghi.
\newblock {On the (In)Security of Mobile Two-Factor Authentication}.
\newblock In {\em International Conference on Financial Cryptography and Data
  Security}, 2014.

\bibitem{psd2-sca}
{European Commission}.
\newblock {Strong customer authentication requirement of PSD2 comes into
  force}.
\newblock
  \url{https://ec.europa.eu/info/publications/190913-safer-payment-services_en},
  2019.

\bibitem{playwright}
Pavel Feldman and open-source contributors.
\newblock Playwright.
\newblock \url{https://playwright.dev/}, 2022.

\bibitem{com.receivesms.online}
{Free Phone Number}.
\newblock {Receive SMS}.
\newblock
  \url{https://play.google.com/store/apps/details?id=com.receivesms.online},
  2021.

\bibitem{twitter-enforces-2fa-employees}
Sergiu Gatlan.
\newblock {Twitter employees required to use security keys after 2020 hack}.
\newblock
  \url{https://www.bleepingcomputer.com/news/security/twitter-employees-required-to-use-security-keys-after-2020-hack/},
  2021.

\bibitem{clave-idp}
{Gobierno de España}.
\newblock {Cl@ve – Electronic Identity for the Administration}.
\newblock \url{https://clave.gob.es/}, 2022.

\bibitem{google-safe-browsing}
{Google Developers}.
\newblock {Google Safe Browsing}.
\newblock 2022.

\bibitem{google-voice}
{Google LLC}.
\newblock {Google Voice}.
\newblock \url{https://voice.google.com/}, 2022.

\bibitem{nistpasswordpolicy}
Paul Grassi, James Fenton, Elaine Newton, Ray Perlner, Andrew Regenscheid,
  William Burr, Justin Richer, Naomi Lefkovitz, Jamie Danker, Yee-Yin Choong,
  Kristen Greene, and Mary Theofanos.
\newblock {Digital Identity Guidelines: Authentication and Lifecycle
  Management}, 2020.

\bibitem{gsma-sim-survey}
{GSM Association}.
\newblock {Access to Mobile Services and Proof of Identity 2020: The Undisputed
  Linkages}.
\newblock
  \url{https://www.gsma.com/mobilefordevelopment/wp-content/uploads/2020/03/Access_to_mobile_services_2020_Singles.pdf},
  2020.

\bibitem{disposable-email-services}
Hang Hu, Peng Peng, and Gang Wang.
\newblock {Characterizing Pixel Tracking through the Lens of Disposable Email
  Services}.
\newblock In {\em 2019 IEEE Symposium on Security and Privacy (SP)}, pages
  365--379, 2019.

\bibitem{cowin}
{India's Ministry of Health and Family Welfare}.
\newblock {CoWIN}.
\newblock \url{https://www.cowin.gov.in/}, 2022.

\bibitem{microsoft-goes-passwordless}
Vasu Jakkal.
\newblock {The passwordless future is here for your Microsoft account}.
\newblock
  \url{https://www.microsoft.com/security/blog/2021/09/15/the-passwordless-future-is-here-for-your-microsoft-account/},
  2021.

\bibitem{wordsegment}
Grant Jenks.
\newblock {WordSegment - Python Word Segmentation}.
\newblock \url{https://grantjenks.com/docs/wordsegment/}, 2018.

\bibitem{com.kwai.video}
{Joyo Technology Pte Ltd}.
\newblock {Kwai – Watch cool \& funny videos}.
\newblock \url{https://play.google.com/store/apps/details?id=com.kwai.video},
  2022.

\bibitem{tranco}
Victor Le~Pochat, Tom Van~Goethem, Samaneh Tajalizadehkhoob, Maciej
  Korczy\'{n}ski, and Wouter Joosen.
\newblock {Tranco: A Research-Oriented Top Sites Ranking Hardened Against
  Manipulation}.
\newblock In {\em Network and Distributed System Security (NDSS) Symposium},
  2019.

\bibitem{insecurity-of-sms-otp}
Zeyu Lei, Yuhong Nan, Yanick Fratantonio, and Antonio Bianchi.
\newblock {On the insecurity of SMS one-time password messages against local
  attackers in modern mobile devices}.
\newblock In {\em Network and Distributed System Security (NDSS) Symposium},
  2021.

\bibitem{pythonsimhash}
Leon.
\newblock Simhash.
\newblock \url{https://github.com/1e0ng/simhash}, 2020.

\bibitem{www2007manku}
Gurmeet~Singh Manku, Arvind Jain, and Anish Das~Sarma.
\newblock {Detecting Near-Duplicates for Web Crawling}.
\newblock In {\em Proceedings of the 16th International Conference on World
  Wide Web}, WWW '07, page 141–150, New York, NY, USA, 2007. Association for
  Computing Machinery.

\bibitem{mdn-fetch-api}
{MDN contributors}.
\newblock {Using the Fetch API}.
\newblock
  \url{https://developer.mozilla.org/en-US/docs/Web/API/Fetch_API/Using_Fetch},
  2022.

\bibitem{com.nhs.online.nhsonline}
{National Health Service}.
\newblock {NHS App}.
\newblock
  \url{https://play.google.com/store/apps/details?id=com.nhs.online.nhsonline},
  2022.

\bibitem{fake-engagement}
David Nevado-Catalán, Sergio Pastrana, Narseo Vallina-Rodriguez, and Juan
  Tapiador.
\newblock {An analysis of fake social media engagement services}.
\newblock {\em {Computers \& Security}}, 124:103013, 2023.

\bibitem{google-enforces-2fa}
Lily~Hay Newman.
\newblock {Google Gets Serious About Two-Factor Authentication. Good!}
\newblock
  \url{https://www.wired.com/story/google-two-factor-authentication-default/},
  2021.

\bibitem{hotstar}
{Novi Digital Entmt.}
\newblock {Disney+ Hotstar}.
\newblock \url{www.hotstar.com/}, 2022.

\bibitem{aadhaar}
{Privacy Intl.}
\newblock {ID systems analysed: Aadhaar}.
\newblock
  \url{privacyinternational.org/case-study/4698/id-systems-analysed-aadhaar},
  2021.

\bibitem{sending-out-an-sms}
Bradley Reaves, Nolen Scaife, Dave Tian, Logan Blue, Patrick Traynor, and
  Kevin~RB Butler.
\newblock {Sending out an SMS: Characterizing the Security of the SMS Ecosystem
  with Public Gateways}.
\newblock In {\em IEEE Symposium on Security and Privacy (S\&P)}, 2016.

\bibitem{sending-out-an-sms-2}
Bradley Reaves, Luis Vargas, Nolen Scaife, Dave Tian, Logan Blue, Patrick
  Traynor, and Kevin~RB Butler.
\newblock {Characterizing the security of the SMS ecosystem with public
  gateways}.
\newblock {\em ACM Transactions on Privacy and Security (TOPS)}, 22(1), 2018.

\bibitem{com.jamtech.marabel.smood}
{Smood.ch}.
\newblock {Smood – Suisse Food Delivery}.
\newblock
  \url{https://play.google.com/store/apps/details?id=com.jamtech.marabel.smood},
  2022.

\bibitem{textnow}
{TextNow, Inc.}
\newblock Textnow.
\newblock \url{https://www.textnow.com/}, 2022.

\bibitem{dialing-back-abuse-on-pva}
Kurt Thomas, Dmytro Iatskiv, Elie Bursztein, Tadek Pietraszek, Chris Grier, and
  Damon McCoy.
\newblock {Dialing Back Abuse on Phone Verified Accounts}.
\newblock In {\em Conf. on Comp. and Comms. Security (CCS)}, 2014.

\bibitem{com.tsoft.moshnumbers}
TSOFT.
\newblock {SMS Receive Phone Numbers}.
\newblock
  \url{https://play.google.com/store/apps/details?id=com.tsoft.moshnumbers},
  2020.

\bibitem{twilio}
Twilio.
\newblock {Communication APIs for SMS, Voice, Video \& Authentication |
  Twilio}.
\newblock \url{https://www.twilio.com/}, 2022.

\bibitem{twilo-sms-traffic-pumping-fraud}
Twilio.
\newblock {SMS Traffic Pumping Fraud}.
\newblock
  \url{https://support.twilio.com/hc/en-us/articles/8360406023067-SMS-Traffic-Pumping-Fraud},
  2022.

\bibitem{unicode-glossary}
{Unicode, Inc.}
\newblock {Glossary of Unicode Terms}.
\newblock \url{unicode.org/glossary/}, 2021.

\bibitem{unicode-normalization-forms}
{Unicode, Inc.}
\newblock {Unicode Normalization Forms}.
\newblock \url{https://unicode.org/reports/tr15/}, 2021.

\bibitem{aadhaar-faq}
{Unique Identification Authority of India}.
\newblock {Aadhaar Myth Busters}.
\newblock
  \url{https://uidai.gov.in/my-aadhaar/about-your-aadhaar/aadhaar-myth-busters.html},
  2022.

\bibitem{twilio-short-code}
Donnie Wang.
\newblock {What is a Short Code?}
\newblock \url{https://www.twilio.com/docs/glossary/what-is-a-short-code},
  2022.

\bibitem{franc}
Titus Wormer.
\newblock {wooorm/franc: Natural language detection}.
\newblock \url{https://github.com/wooorm/franc}, 2021.

\end{thebibliography}

\end{document}